\documentstyle[12pt]{article}

\begin{document}

\thispagestyle{empty}

\hfill \parbox{45mm}{{UTF-368/96} \par Jan 1996} 

\vspace*{15mm}

\begin{center}
{\LARGE Role of a "Local" Cosmological Constant}

\smallskip

{\LARGE in Euclidean Quantum Gravity.}
        
\vspace{22mm}

{\large Giovanni Modanese}%
\footnote{e-mail: modanese@science.unitn.it}

\medskip

{\em I.N.F.N. -- Gruppo Collegato di Trento \par
Dipartimento di Fisica dell'Universita' \par
I-38050 POVO (TN) - Italy}

\bigskip \bigskip

\medskip

\end{center}

\vspace*{10mm}

\begin{abstract}
In 4D non-perturbative Regge calculus a positive value 
of the effective cosmological constant characterizes the
collapsed phase of the system. If a local term of the form
$S'=\sum_{h \epsilon \{h_1,h_2,...\} } \lambda_h V_h$ is added to
the gravitational action, where $\{h_1,h_2,...\}$ is a subset of
the hinges and $\{\lambda_h\}$ are positive constants, one expects
that the volumes $V_{h_1}$, $V_{h_2}$, ... tend to collapse 
and that the excitations of the lattice propagating 
through the hinges $\{h_1,h_2,...\}$
are damped. We study the continuum analogue of this effect.
The additional term $S'$ may represent the coupling of the
gravitational field to an external Bose condensate.

\medskip
\noindent
04.20.-q Classical general relativity.

\noindent
04.60.-m Quantum gravity.

\bigskip 

\end{abstract}

Since the first perturbative formulations of quantum gravity 
it was realized that the addition of a cosmological term 
$\frac{\Lambda}{8\pi G} \int d^4 x \, \sqrt{g(x) } $ to 
the pure Einstein action gives the graviton a mass, which is 
positive if $\Lambda < 0$ and negative -- that means, the theory 
becomes unstable -- if $\Lambda > 0$ \cite{veltman}.

Nevertheless, whilst the vacuum fluctuations of the quantum 
fields should in principle produce a very large value of the 
cosmological constant, there is no observable hint of it. 
Namely, neither the Newtonian potential shows any finite range 
up to solar system distances, nor pure gravity
exhibits any instability in the weak field case;
finally the observations on cosmological scale set an upper limit
on $|\Lambda|$ as small as $|\Lambda| < 10^{-120} G^{-1}$
in natural units. This discrepancy is known as the "cosmological
constant problem" and several possible escapes have been 
suggested \cite{weinberg}. Treating the problem at a
fundamental level requires a non-perturbative approach, since one
should be able to explain why the actual large-scale geometry of 
spacetime is flat just from dynamic considerations. It is therefore
not surprising that a definitive and generally accepted solution 
of the paradox of the cosmological constant is still remote. 
We shall not make any attempt to a new explanation here.

We shall consider 4D pure gravity in the Euclidean approach,
with special reference to quantum Regge calculus \cite{h1}. In this 
model the results of the numerical non-perturbative simulations
compose the following picture of the behaviour of the cosmological
constant: while its "bare" value $\lambda$ is generally nonzero,
the effective value $\Lambda$ depends on the energy scale $\mu$
and vanishes at large distances like $|\Lambda| \sim G^{-1} 
(\mu l_0)^\gamma$, where $l_0$ is the lattice spacing and 
$\gamma$ a critical exponent. This means that the quantum 
geometry fluctuates on small scales, but reproduces 
flat space at macroscopic distances. The sign of $\Lambda$ is 
negative, thus the flat limit is well defined.

Our aim is to study the interaction of
this gravitational system with a particular external source,
namely a Bose condensate described by a scalar field
$\phi(x)=\phi_0(x) + \tilde{\phi}(x)$. We assume that the
vacuum density $\phi_0(x)$ is forced from the outside to a
certain value, as it can happen for instance in a superconductor
subjected to external electromagnetic fields. The positivity of the
Euclidean action of $\phi$ ensures that the terms $[\partial^\mu
\phi_0(x)]^* [\partial_\mu \phi_0(x)]$ and $m^2_\phi |\phi_0(x)|^2$
act like positive cosmological contributions to the
gravitational action, possibly inducing local gravitational 
instabilities.

The structure of the paper is the following. In Section 1 we recall
the main results of Euclidean 4D quantum Regge calculus concerning 
the effective cosmological constant. We mention its scale behaviour and
two possible interpretations of the lattice spacing $l_0$. 
In Section 2 we consider in the continuum theory, at distances
much larger than Planck scale, a weak Euclidean gravitational
field $g_{\mu \nu}(x)=\delta_{\mu \nu} +
\kappa h_{\mu \nu}(x)$, whose action includes an infinitesimal
effective cosmological term with $\Lambda <0$. Such a theory can
be regarded as the continuum limit of the lattice theory described
in Section 1. We then couple minimally $g_{\mu \nu}(x)$
to the mentioned scalar field $\phi(x)$, with vacuum
density $\phi_0(x)$ determined from the outside, and write in
detail the various terms of the action. It turns out that the
infinitesimal graviton mass $m^2_g \propto |\Lambda|$ receives a
local negative contribution, which we denote by $-\mu^2(x)$.
This means, as we show explicitly for the simpler
case of an almost-massless scalar field $\chi$, that if there
exist some four-dimensional regions in which $\mu^2(x)>m^2_\chi$,
it is possible to find field configurations $\chi(x)$ which
make the Euclidean action unbounded from below. The field
tends to develope singularities in those regions, or -- if a
cut-off mechanism comes into play -- it tends to assume constant
extremal values which are independent from those
in the neighboring regions. We suggest that these "constraints"
should be inserted in the equation for the propagator of the
field.

In Section 3 we remind the formula which gives the static
potential energy in Euclidean quantum gravity as a functional average 
\cite{m,h2}. We illustrate its statistical meaning in the case
of a weak field through an analogy with a simple 2D Ising
model, which we also treat numerically with an elementary
simulation. We show numerically that the insertion of local
supplementary constraints due to an external field (in analogy
to the singularities of the continuum case)
damps in a sensible way the spin-spin correlations, and thus
the interaction they represent in the model.

\section{Regge calculus.}

In the last years discretized quantum gravity on the Regge lattice 
has led, through the Montecarlo numerical simulations of 
Hamber and Williams, to a better comprehension of the non-perturbative 
behaviour of the Euclidean gravitational field in four dimensions. 
Some of the features which emerge from their results \cite{h1} depend 
on the specific model, but several others are 
quite general. In this approach the physical quantities are extracted 
from functional averages. The partition function is written as
\begin{equation}
  Z=\int_{Geometries} d[g] \, e^{-\hbar^{-1} S[g]}
\label{z}
\end{equation}
where the geometries are described by piecewise flat simplicial 
manifolds. The integral (i.e.\ the Montecarlo sampling) runs over 
the lengths of the links which define a field configuration. The 
action in (\ref{z}) has the form
\begin{equation}
  S = \int d^4 x \, \sqrt{g} \left( \lambda - kR +
  \frac{1}{4} a R_{\mu \nu \rho \sigma} R^{\mu \nu \rho \sigma}
  \right)
\label{action}
\end{equation}
or, in discretized version,
\begin{equation}
  S[l] = \sum_{hinges \ h} V_h \left[\lambda - k \frac{A_h \delta_h}{V_h}
  + a \frac{A_h^2 \delta_h^2}{V_h^2} \right]
\label{discr}
\end{equation}
where $A_h$ is the area of an hinge, $V_h$ its volume and $\delta_h$
the defect angle (see the original papers for the definitions
of lattice quantities and the functional measure).

In the following we shall set $\hbar=1$. The constants $k$ and 
$\lambda$ are related, as "bare" quantities, to the 
Newton constant $G$ and to the cosmological constant $\Lambda$: $k$
corresponds to $1/8\pi G$ and $\lambda$ to $\Lambda/8\pi G$. It is
important, however, to keep distinct the physical values $G$ and
$\Lambda$ from $k$ and $\lambda$. The latter are entered as parameters
at the beginning, and then a second order transition point for the
statistical system described by $Z$ is found by Montecarlo simulation.
Actually, there is a line of transition, since one can also vary the
adimensional parameter $a$, which does not have a macroscopic
counterpart. On this line in the parameter space the theory admits
a continuum limit. Unlike in perturbation theory, where a flat
background is introduced by hand, here the flat space appears 
dynamically; namely, the average value of the curvature is found to
vanish on the transition line, which separates a "smooth phase",
with small negative curvature, from a "rough", unphysical phase, with
large positive curvature. In this way the effective, large scale
cosmological constant
\begin{equation}
  \Lambda = \frac{\left< \int \sqrt{g} \, R \right>}
  {\left< \int \sqrt{g} \right>}
\label{cceff}
\end{equation}
vanishes in the continuum quantum theory. 

More precisely, the dependence of the effective (or "running")
cosmological constant $\Lambda_{eff}$ on the scale is the following.
If we compute (\ref{cceff}) on small volumes, the curvature fluctuates
wildly. At larger distances the average curvature decreases, 
because the fluctuations tend to average out. If $\mu$ denotes the
energy scale, close to the critical point the adimensional quantity
$|\Lambda|G$ behaves like
\begin{equation}
  (|\Lambda|G)(\mu) \sim (l_0 \mu)^\gamma ,
\label{scale}
\end{equation}
where $l_0$ is the average spacing of the dynamical lattice and
$\gamma \sim 1.56$ in the first simulations. The sign of $\Lambda$
is negative, as mentioned. Since the $\mu$-dependence of $G$ is quite 
weak, the effective cosmological constant decreases approximately
like a power law as the length scale grows.

Eq.\ (\ref{scale}) admits two different physical interpretations, 
depending on the role we attribute to the average lattice spacing $l_0=
\sqrt{\langle l^2 \rangle}$. In the usual lattice theories, $l_0$ is sent 
to zero in order to obtain the continuum limit. In this first interpretation, 
eq.\ (\ref{scale}) shows the way the lattice theory reproduces flat
space in the physical limit. No real physical meaning is assigned
to the effective $\Lambda$ which, at a fixed scale $\mu$,
is simply proportional to a positive power of the regulator.

On the other hand, we can believe that in quantum gravity $l_0$
has an intrinsic minimum value of the order of Planck length
$L_{Planck} \sim 10^{-33} \ cm$. This hypotesis arises independently 
from several operational models (for a review see \cite{planck}) or 
from more complex quantum theories (see for instance \cite{rovsmo}
and references). In the framework of Regge calculus, it is possible to
fix $l_0$ by imposing that the effective Newton constant $G$
computed non-perturbatively \cite{h2} is of the same magnitude order of
the observed value; in this way one finds one more time that
$l_0$ has to be of the order of Planck length \cite{graviton}.

In this second case, that is $l_0 \sim L_{Planck}$, the 
interpretation of eq.\ (\ref{scale}) is different: it means that 
the effective cosmological constant tends to zero on large scale,
while it is nonvanishing, in principle, on small scale (here
and in the following we mean by "large" scale the laboratory
or atomic scale, and by "small" scale the Planck scale). This
interpretation does not necessarily have observable physical 
consequences, since in fact $\Lambda$ could be far too small.
Namely, the exponent gamma has been computed only for small lattices; 
an evaluation for lattices of "macroscopic" size is of course
technically impossible, and thus only the experiments could tell us
whether the law (\ref{scale}) keeps true for large distances, and with
which exponent. The fact that on astronomical scale we have
$|\Lambda|G < 10^{-120}$ for $\mu \sim 10^{-30} cm^{-1}$
constrains $\gamma$ to be approximately larger than 2. But the vanishing 
could be much more rapid, so that we could disregard $\Lambda$ at
any physically relevant scale.

Another property of Regge calculus which shows an
intrinsic feature of quantum gravity is the instability of the
"rough" phase with positive average curvature, i.e. with positive 
effective cosmological constant. This phase does not admit any 
continuum limit. Its fractal dimension is small,
which denotes that the geometry is collapsed.

The non-perturbative instability properties of the phase with 
positive average curvature extend the validity of known 
considerations based on perturbation theory. Namely, in the weak 
field approximation a positive cosmological
constant in the gravitational lagrangian produces a negative
mass for the graviton (compare \cite{veltman} and the next Section). 
Instabilities arise in the perturbative theory on a De Sitter 
background too \cite{tw,z}. On the contrary, a small negative 
cosmological constant generally does not imply any instability, but 
a small mass $m_g$ for the graviton, of the order of $\Lambda^{1/2}$
in natural units.

In conclusion, at a scale large enough with respect to the 
lattice scale it is possible to regard the discretized gravitational 
functional integral (\ref{z}) as describing an almost flat 
mean field plus fluctuations. If the fixed point of the lattice theory is
approached from the physical, smooth phase, the effective 
cosmological constant $\Lambda$ (i.e.\ the average of the scalar 
curvature $R$) is very small and negative, and the large scale
fluctuations of $R$ are small too. The system is stable,
because the field configurations with larger volume, in which
the links are as stretched as possible, are preferred to the
collapsed configurations, since the Euclidean action depends on the    
volume like $S \sim \frac{\Lambda}{8\pi G} \int d^4x \,     
\sqrt{g} \sim \frac{\Lambda}{8\pi G} \sum_h V_h$.
(On the contrary, a positive value of $\Lambda$ would favour the 
collapsed configurations with smaller volume.) In this picture
$\Lambda$ can be regarded either as a purely formal regulator, which 
goes to zero in the physical limit $l_0 \to 0$, or as
a physical quantity, though possibly extremely small (in the
second interpretation, in which $l_0 \sim l_{Planck}$).

Keeping these properties in mind, it is interesting to 
consider the case in which the coupling of the Euclidean gravitational 
field with an external source gives in some four-dimensional regions a 
positive contribution to the effective cosmological constant.
In the next Section we shall analyse this phenomenon in the
continuum case, that is, on large scale (in the meaning of "large
scale" we precised above).

On the lattice, such a coupling would correspond in the action 
(\ref{discr}) to an additional term of the form
\begin{equation}
  \sum_{\{h_1, h_2 ...\} } \lambda_h V_h
\end{equation}
where $\{h_1, h_2 ...\}$ is a subset of the hinges and $\lambda_h$
are fixed positive constants.

We expect that when the Montecarlo algorithm chooses 
for the random variation a link which ends in a hinge $h_i \epsilon 
\{h_1, h_2 ...\}$, the favoured variation will be that for
which the volume $V_{h_i}$ decreases. Thus the volumes 
$\{ V_{h_1}, \ V_{h_2}, ...\}$ will tend to collapse and the
lattice excitations propagating through the hinges
$\{h_1, h_2 ...\}$ will be damped.

\section{Continuum case.}

We have seen that the results of quantum Regge calculus can be
interpreted as leading at distances large compared to $L_{Planck}$
to an effective Euclidean action for pure gravity of the form
\begin{equation}
  S_{eff}=\int d^4x \, \sqrt{g(x)} \left[
  \frac{\Lambda}{8\pi G} - \frac{1}{8\pi G} R(x) \right]
\label{aga}
\end{equation}
where the curvature fluctuations around flat space are small
and the effective cosmological constant $\Lambda$ is negative
and very small too. (As we saw in the previous Section, $\Lambda$
is scale-dependent; we suppose here to stay at some fixed 
scale.) From the geometrical point of view, the
small negative value of $\Lambda$ stabilizes the system, preventing 
it from falling into small-volume, collapsed configurations. 

In the naive perturbation theory around the flat
background the $\Lambda$-term represents a small mass for the graviton
\footnote{As remarked in \cite{veltman}, this widespread
belief is not rigorously true. It has also been proved
\cite{vdam} that all theories of a massive tensor field
in Minkowski space which satisfy the usual QFT postulates are
incompatible with General Relativity in the limit of vanishing
mass. In our reasoning it is not essential to regard the
cosmological term as a graviton mass term, but for simplicity
we stick to this terminology.}.
Namely, setting $g_{\mu \nu}(x)=\delta_{\mu \nu}
+ \kappa \tilde{h}_{\mu \nu}(x)$, with $\kappa=\sqrt{8\pi G}$,
the determinant $g$ of $g_{\mu \nu}$ can be expanded as 
\begin{equation}
  g=1+\kappa \tilde{h}^{(1)}+\kappa^2 \tilde{h}^{(2)} + ... \ ,
\end{equation}
where $\tilde{h}^{(1)}$, $\tilde{h}^{(2)}$, ... denote
expressions which are linear, quadratic etc.\ in 
$\tilde{h}_{\mu \nu}$. The linear "tadpole" term $\tilde{h}^{(1)}$
is usually disregarded, since it is proportional to the trace
$\tilde{h}^{\mu}_{\mu}$, which vanishes on physical states.
The term $\tilde{h}^{(2)}$ takes the form of a graviton mass term,
such that the mass is positive when $\Lambda < 0$ (compare 
\cite{veltman}). We thus have
\begin{eqnarray}
  & S_{eff} = \int d^4x \Bigl\{ & \left[ m^2_g \tilde{h}^{(2)}(x)
  - \tilde{R}^{(2)}(x) \right] + \Lambda \left[ \kappa \tilde{h}^{(3)}(x)+
  \kappa^2 \tilde{h}^{(4)}(x) + ... \right] + \nonumber \\
  & & \ \ - \left[ \kappa \tilde{R}^{(3)}(x)+ \kappa^2 \tilde{R}^{(4)}(x) + 
  ... \right] \Bigr\} ,
\label{quadr}
\end{eqnarray}
where $m_g$ is proportional to $\Lambda^{1/2}$ and $\tilde{R}^{(2)}$, 
$\tilde{R}^{(3)}$ ... denote the parts of $R$ which are quadratic, 
cubic etc.\ in $\tilde{h}_{\mu \nu}$.

The first bracket contains the quadratic part of the action.
The third bracket contains the familiar self-interaction vertices
of the graviton, involving respectively 1, 2, ... derivatives of
$h_{\mu \nu}$. The second bracket contains self-interaction
vertices which are peculiar of the theory with $\Lambda \neq 0$
and do not involve derivatives. We are however not interested
in the self-interaction vertices of $h_{\mu \nu}$ in the following.

Now we would like to consider the interaction of $\tilde{h}$
with a scalar field $\phi$ having non-vanishing
vacuum expectation value $\phi_0$. We suppose $\phi_0$ to be
spacetime dependent and denote $\phi(x)=\phi_0(x)+\tilde{\phi}(x)$;
$\phi_0(x)$ is regarded as a quantity determined from the outside, that is,
as a source term. In this way, the field $\phi(x)$ describes a Bose 
condensate with ground state density $\phi_0(x)$ fixed by external 
conditions (like, for instance, the Cooper pairs density in certain
superconductors under external e.m.\ field). 

The total action is
\begin{eqnarray}
  & S = S_{eff} + \int d^4x \, \sqrt{g(x)} & \Bigl\{
  \partial_\mu \left[ \phi_0(x) + \tilde{\phi}(x) \right]^*
  \partial_\nu \left[ \phi_0(x) + \tilde{\phi}(x) \right]
  g^{\mu \nu}(x) + \frac{1}{2} m_\phi^2 |\phi_0(x)|^2 \nonumber \\
  & &  + \frac{1}{2} m_\phi^2
  \left[ \phi_0^*(x) \tilde{\phi}(x) + \phi_0(x) \tilde{\phi}^*(x) \right]
  + \frac{1}{2} m^2 |\tilde{\phi}(x)|^2 \Bigr\}
\end{eqnarray}

We can rewrite it as
\begin{equation}
  S=\int d^4x \, \sqrt{g(x) } \left\{ \left[
  \frac{\Lambda}{8 \pi G} + \frac{1}{2} \mu^2(x) \right]
  - \frac{1}{8 \pi G} R(x) \right\} + S_1 + S_2 ,
\end{equation}
where
\begin{eqnarray}
  \frac{1}{2} \mu^2(x) & = & \frac{1}{2}
  [\partial_\mu \phi_0^*(x)] [\partial^\mu \phi_0(x)] +
  \frac{1}{2} m^2_\phi |\phi_0(x)|^2 ; \\
  S_1 & = & \frac{1}{2} \int d^4x \, \sqrt{g(x) }
  \, \partial_\mu \phi_0^*(x) \partial_\nu \phi_0(x) \kappa
  \tilde{h}^{\mu \nu}(x) ; \\
  S_2 & = & \frac{1}{2} \int d^4x \, \sqrt{g(x) }
  \Bigl\{ m^2_\phi |\tilde{\phi}(x)|^2 + m^2_\phi
  \left[ \phi_0^*(x) \tilde{\phi}(x) + \phi_0(x) \tilde{\phi}^*(x)
  \right] + \nonumber \\
  & & + \left[ \partial_\mu \tilde{\phi}^*(x) 
  \partial_\nu \tilde{\phi}(x) + \partial_\mu \phi_0^*(x)
  \partial_\nu \tilde{\phi}(x) + \partial_\mu \tilde{\phi}^*(x)
  \partial_\nu \phi_0(x) \right] g^{\mu \nu}(x) \Bigr\}
\end{eqnarray}
  
For brevity, we have not expanded here $\sqrt{g}$ and $R$ like 
in eq.\ (\ref{quadr}). Let us first look at the terms $S_1$ and $S_2$.
The term $S_1$ describes a process in which gravitons are produced
by the source $\phi_0(x)$. The term $S_2$ contains the free
action of the field $\tilde{\phi}(x)$, which describes the excitations
of the condensate, and several vertices in which the graviton field 
$\tilde{h}_{\mu \nu}(x)$ and $\tilde{\phi}(x)$ interact between 
themselves and possibly with the source. All these interactions
are not of special interest here and are generally very weak, 
due to the smallness of the coupling $\kappa$. The relevant point
is that the purely gravitational cosmological term 
$\frac{\Lambda}{8 \pi G}$ receives a "local" positive contribution 
$\frac{1}{2}\mu^2(x)$ which depends on the fixed external source $\phi_0(x)$. 
According to eq.s (\ref{quadr}), (\ref{aga}) and to our discussion of the
sign of $\Lambda$, this amounts to a negative mass contribution
and could lead to instabilities.

Let us study the effect of such a local negative mass contribution in 
the simpler case of a scalar field in flat space. We consider a scalar 
field $\chi$ with very small mass $m_\chi$ and add to its free 
Euclidean action a source term of the form $-\frac{1}{2} \mu^2(x) 
\chi^2(x)$, which represents a "localized negative mass". 
The action becomes
\begin{equation}
  S_\chi = \frac{1}{2} \int d^4x \left\{
  [\partial^\mu \chi(x)] [\partial_\mu \chi(x)] +
  m_\chi^2 \, \chi^2(x) - \mu^2(x) \chi^2(x) \right\}
\end{equation}

To fix the ideas, let us suppose that $\mu^2(x)$ is different 
from zero in certain four-dimensional regions $\Omega_i$, where it 
takes the constant values $\mu_i$. Outside these regions let
$\mu^2(x)$ go rapidly to zero.
The solution of the classical field equation for $\chi$ is obtained
by minimizing the action. In the absence of the source term we
would have of course $\chi(x)=const.=0$, because this minimizes at
the same time the gradient term $\frac{1}{2}[\partial^\mu \chi(x) 
\partial_\mu \chi(x)]$ and the mass term $\frac{1}{2} m_\chi^2 
\, \chi^2(x)$ (both positive-defined).

In the presence of the source term $-\frac{1}{2} \mu^2(x) \chi^2(x)$, the
action decreases when $\chi^2$ becomes large within the regions $\Omega_i$.
The growth of $\chi$ in these regions is limited only by the
fact that outside them $\chi$ must go to zero (due to the term $m_\chi^2$)
and that the gradient in the transition region cannot in turn be too
large.
Let us suppose for instance that there is only one region $\Omega_0$,
with the shape of a 4-sphere of radius $r_0$. Let $\mu(x)$ take the value
$\mu_0$ inside $\Omega_0$ and zero outside. We consider for $\chi$
the trial function $\chi(r)=\chi_0 \, f(r)$. The action becomes (we admit
spherical symmetry)
\begin{equation}
  S_\chi = \pi^2 \chi_0^2 \int_0^\infty dr \, r^3 \left\{ [f'(r)]^2
  + m_\chi^2 [f(r)]^2 - \mu_0^2 \theta(r-r_0) [f(r)]^2 \right\}
\end{equation}
We see that if the integral is positive, the value of $\chi_0$ which
minimizes the action is still $\chi_0=0$. On the contrary, if the
integral is negative, the action is not bounded from below as
$\chi_0$ grows.

We choose the following explicit form of $f(r)$: for 
$r<r_0$ let $f(r)=1$, i.e., $\chi(r)=\chi_0$; for $r>r_0$ let be
$f(r)=\exp [-(r-r_0)/\delta]$. We thus have
\begin{eqnarray}
  S_\chi & = & \pi^2 \chi_0^2 \left\{ \left( \frac{1}{\delta^2}
  + m_\chi^2 \right) \int_{r_0}^\infty dr \, r^3 \,
  e^{-2(r-r_0)/\delta} - \frac{1}{4} (\mu_0^2 - m_\chi^2) r_0^4 \right\} \\
  & = & \pi^2 \chi_0^2 \left\{ \left( \frac{1}{\delta^2}
  + m_\chi^2 \right) \left( \frac{3}{8} \delta^4 +
  \frac{3}{4} r_0 \delta^3 + \frac{3}{4} r_0^2 \delta^2
  + \frac{1}{2} r_0^3 \delta \right) - 
  \frac{1}{4} (\mu_0^2 - m_\chi^2) r_0^4 \right\}
  \label{boh}
\end{eqnarray}
It is easy to check that for suitable choices of the parameters
the expression within the braces in eq.\ (\ref{boh}) turns out to be 
negative. For instance, if $m_\chi$ can be disregarded with respect 
to $\mu_0$ and $\delta \sim r_0$, the expression is negative 
provided $\mu_0 \gg \delta^{-1}$. If $\delta \ll r_0$ the
expression is negative provided $\mu_0 \gg r_0^{-1}$, etc.
Thus the system is unstable. For the graviton the instability is
even worse, because the kinetic term $\tilde{R}^{(2)}$
is not positive-definite.

Physically, we might of course invoke some additional "regularizing" 
process which comes into play for large values of $\chi_0$ and cuts the
action. Thus the value of $\chi^2(x)$ inside the
region is forced by the source to a certain maximum, and
this will affect the propagation of the field. The differential 
equation for the propagator of $\chi(x)$ in the presence of the source
must now satisfy additional boundary conditions on each region $\Omega_i$.
This amounts to a very complex mathematical problem; the physical
consequence will be a "damping" of the correlations of $\chi(x)$.
One can check this numerically in some model; we shall do this shortly
in the next Section, referring to a toy bidimensional spin model.

\section{The formula for the static potential energy.}

In this Section we remind the formula which gives the static
potential energy in Euclidean quantum gravity as a functional average 
\cite{m,h2}. We illustrate its statistical meaning in the case
of a weak field through an analogy with a simple 2D Ising
model, which we also treat numerically with an elementary
simulation. We show numerically that the insertion of local
supplementary constraints due to an external field (in analogy
to the singularities of the continuum case)
damps in a sensible way the spin-spin correlations, and thus
the interaction they represent in the model.

Let us first consider, in Euclidean quantum field theory, a scalar field
$\Phi(x)$ with action $S_0[\Phi]=\int d^4 x \,L(\Phi(x))$. In the presence 
of the external source $J(x)$, the ground state energy of the system can
be expressed as
\begin{eqnarray}
  E & = & \lim_{T \to \infty} -\frac{\hbar}{T} \log
  \frac{\int d[\Phi] \exp \left\{ - \hbar^{-1} \left[ \int
  d^4 x \, L(\Phi(x) ) + \int d^4 x \, \Phi(x) J(x) \right] \right\} }
  { \int d[\Phi] \exp \left\{ - \hbar^{-1} \int d^4 x \, L(\Phi(x) )
  \right\} } \nonumber \\
  & = & \lim_{T \to \infty} -\frac{\hbar}{T} \log
  \left< \exp \left\{ - \hbar^{-1} \int d^4 x \, \Phi(x) J(x) 
  \right\} \right> ,
\label{aa}
\end{eqnarray}
where it is assumed that the source vanishes outside the interval 
$(-T/2,T/2)$ and that the coupling between $J$ and $\Phi$ 
is linear. More generally, a formula similar to (\ref{aa}) holds when we 
are dealing with more fields $\Phi_A$ and corresponding sources $J^A$, 
and when the coupling between fields and sources is not linear.

As an useful application of eq.\ (\ref{aa}), we can write the interaction 
energy $E(L)$ of two static pointlike sources of the field, kept at 
a fixed distance $L$. We just need to insert the suitable
expression for $J$. The trajectories of the two sources are in flat space
\begin{equation}
  x^\mu_1(t_1) = (t_1, \, 0, \, 0, \, 0); \qquad
  x^\mu_2(t_2) = (t_2, \, L, \, 0, \, 0) .
\label{tra}
\end{equation}

In ordinary gauge theories we may re-obtain in this way the
Wilson formula for the static quark-antiquark potential.
In quantum gravity we are led to the following equation 
for the static potential \cite{m}
\begin{eqnarray}
  E(L) & = & \lim_{T \to \infty} - \frac{\hbar}{T}
  \log \frac{\int d[g] \, \exp \left\{ - \hbar^{-1} \left[
  S[g] + \sum_{i=1,2} m_i \int_{-\frac{T}{2}}^{\frac{T}{2}} dt \,
  \sqrt{g_{\mu \nu}[x_i(t)] \dot{x}_i^\mu(t) \dot{x}_i^\nu(t)}
  \right] \right\}}{\int d[g] \, \exp \left\{ - \hbar^{-1} S[g] \right\} }
  \nonumber \\
  & & \label{ciao} \\
  & \equiv & \lim_{T \to \infty} - \frac{\hbar}{T} \log 
  \left< \exp \left\{ - \hbar^{-1} \sum_{i=1,2} m_i 
  \int_{-\frac{T}{2}}^{\frac{T}{2}} ds_i \right\} \right>_S
\label{bella}
\end{eqnarray}
where $S$ is the euclidean action 
\footnote{Notice that this formula applies also to two masses which are not
pointlike ("pointlike particle" is actually an ill-defined concept in 
General Relativity), provided we can disregard the internal degrees of
freedom. Namely, their action is still equal to $\sum_i \int ds_i$, where 
the integrals are taken along the trajectories of the centers of mass.}.

In eq.s (\ref{ciao}), (\ref{bella}) the lines $x_1(t)$ and $x_2(t)$
must be parallel with respect
to the dynamic metric $g_{\mu \nu}$ and thus they should in
principle be re-traced for each field configuration of the
functional integral. In practice, it is extremely difficult to 
compute a functional integral defined in such a formal way. 
Let us then limit ourselves to consider weak fluctuations 
of the gravitational field about flat 
space. The trajectories of the sources may be defined with respect to the 
background metric like in (\ref{tra}). It is straightforward to 
reproduce in this way the Newton potential energy \cite{m,h2}; higher 
order corrections have been computed too \cite{muz}. 
The geodesic distance between the trajectories $x_1(t)$ and $x_2(t)$
is now equal to $L$ only on the average; in fact such an approximation 
is not without physical meaning, since in any realistic source the 
fixed distance at which the two masses are kept can only be an 
average value. Also in the non-perturbative evaluations of eq.\
(\ref{bella}) in quantum Regge gravity \cite{h2}, the distance $L$ 
is evaluated {\it a posteriori} as the mean value of the geodesic distance
on all configurations. 

Let us consider the almost-flat metric $g_{\mu \nu}=\delta_{\mu \nu}
+\eta_{\mu \nu}$ in a fixed gauge. We obtain
\begin{equation}
  \int_{-T/2}^{T/2} ds_i \simeq \int_{-T/2}^{T/2} dt_i \,
  \sqrt{1+h_{11}[x(t_i)]} \simeq T + \frac{1}{2}
  \int_{-T/2}^{T/2} dt_i \, h_{11}[x(t_i)]
\label{acca}
\end{equation}
and we see that to a first approximation the fluctuations of $h$
have the effect of making each line "shorter" or "longer".
Let us call $\alpha[h]$ the (gauge-invariant) difference between the 
lenghts of the two lines in a field configuration $h$, and assume 
for simplicity that the masses of the two sources are equal: $m_1=m_2=m$.
We may expand the exponential in eq.\ (\ref{bella}), finding 
(note that $\langle \alpha[h] \rangle$ obviously vanishes by symmetry)
\begin{equation}
  E = 2m + \lim_{T \to \infty} - \frac{\hbar}{T} \log
  \left< \exp \{ - \hbar^{-1} m \alpha[g] \} \right>
  = 2m + \lim_{T \to \infty} - \frac{m^2}{2\hbar T}
  \left< \alpha^2[g] \right> + ...
\label{stat}
\end{equation}
This equation exibits an interesting relation between the vacuum
fluctuations of the geometry and the static gravitational potential.
To illustrate better its "statistical" meaning
we would like now to introduce a toy analogy with the 2D Ising model. 

Let us consider a planar
spin system with periodic boundary conditions and the local
coupling $H=-J\sum_{i,j} s_{ij} s_{i'j'}$ ($J>0; \ (i',j')$
neighbours of $(i,j)$). Let us then consider two columns $j_1$ 
and $j_2$ at a distance of $L$ lattice spacings (see fig.\ 1).

We can regard this system as the analogue of a discretized
configuration of a 4D gravitational field on the plane between the
two parallel lines of eq.\ (\ref{tra}). The spin variables $\pm 1$ 
represent fluctuations of the metric. At the transition temperature, 
the fluctuations of the spin variables along the two lines are 
correlated, approximately like $1/L$.

Going back to eq.\ (\ref{acca}) and making the correspondence
$s \leftrightarrow h_{11}$, we see that the analogue of
$\alpha[g]$ is the difference between spin sums taken along the
columns $j_1$ and $j_2$. The analogue of eq.\ (\ref{stat}) is
\begin{eqnarray}
  \langle \alpha^2_{Ising} \rangle & = & \left< \left( \sum_i s_{ij_1}
  - \sum_i s_{ij_2} \right)^2 \right> = \nonumber \\
  & = & \left< \left( \sum_i s_{ij_1} \right)^2 +
  \left( \sum_i s_{ij_2} \right)^2 -
  2 \sum_i s_{ij_1} \sum_k s_{kj_2} \right>
\end{eqnarray}

We are interested only in the term which depends on the distance $L$
between the two columns, that is, the product term
\begin{equation}
  p_{12} = \left< \sum_i s_{ij_1} \sum_k s_{kj_2} \right>. 
\end{equation}

\newpage

\begin{verbatim}

   + - - + -  +  + - - +  -  - + - + + - 
   + + + - -  +  + - + -  +  - - + - + + 
   - - + + -  +  - + + +  -  - + - + + - 
   + - + + -  -  + + + -  +  - + + - + + 
   - + - - +  -  + + - +  +  + - + - + + 
   + + + - -  +  + - + -  +  - - + - + + 
   - + - - +  +  - + - -  +  + - - + - - 
   - - + + -  +  - + + +  -  + + - + + - 
   + + - + -  +  + - - +  -  - + - + + - 
   - + - - +  -  + + - +  +  + - + - + + 
   - - + + -  +  - + + +  -  - + - + + - 
              |           |
             j1          j2

               <----L---->

   Fig. 1 - Spin sums taken along two columns of a 2D Ising system.
\end{verbatim}

A numerical simulation with a simple $10 \times 12$ system has given, 
as expected, the following results at the critical temperature:
\begin{verbatim}
  L (lattice spacings)           p_{12}
  ===============================================
            2             1.15   +/-   0.13
            3             0.74   +/-   0.09
            4             0.48   +/-   0.09
            5             0.42   +/-   0.08
            6             0.41   +/-   0.08

\end{verbatim}

We see that the $L^{-1}$ law is approximately verified also for
the correlation between the spin sums taken along the two
columns. Only for $L=6$ there is a deviation, which can be
explained as due to the periodic boundary conditions.

We observe that, due to the nature of the spin system, the correlation
between two spin sums along $j_1$ and $j_2$ is necessarily positive.
Thus, the part of $\langle \alpha^2_{Ising}
\rangle$ which depends on the distance $L$ is negative. In perturbative 
quantum gravity one finds instead that the correlation 
$\langle h_{11}(x)h_{11}(y) \rangle$ is always negative, which
leads to the correct negative sign for the potential energy in
eq.\ (\ref{stat}).

\begin{verbatim}

  0 x 0 0 0 0 x 0 0 0 0 0            0 x 0 0 0 0 x 0 0 0 0 0
  0 x 0 0 0 0 x 0 0 0 0 0            0 x 0 * 0 0 x 0 0 0 0 0
  0 x 0 * 0 0 x 0 0 0 0 0            0 x 0 0 0 0 x 0 0 0 0 0
  0 x 0 0 0 0 x 0 0 0 0 0            0 x 0 0 * 0 x 0 0 0 0 0
  0 x 0 0 * 0 x 0 0 0 0 0            0 x 0 0 0 0 x 0 0 0 0 0
  0 x 0 0 0 0 x 0 0 0 0 0            0 x 0 0 * 0 x 0 0 0 0 0
  0 x 0 * 0 0 x 0 0 0 0 0            0 x 0 0 0 0 x 0 0 0 0 0
  0 x 0 0 0 0 x 0 0 0 0 0            0 x 0 * 0 0 x 0 0 0 0 0
    |         |                        |         |
    j1        j2                       j1        j2

            (a)                                 (b)

  Fig. 2 - Insertions of singular points "*" between the columns
           j1 and j2.

\end{verbatim}

We now introduce some supplementary conditions, in order to
simulate the case in which the spin variables assume on certain
sites a fixed value. This could be due, like in the gravitational
case to which we are interested, to the localized action of an 
external field. In the spin model we may imagine that an external
magnetic field localized on certain sites forces spin-flips.
With reference to Fig.\ 2, let us suppose that the spin-flips
occurr on the sites marked with a star and placed between the
two columns $j_1$ and $j_2$ whose correlation we are measuring.
(The two columns are denoted by "x", while all the remaining
sites are denoted by "0".) To prevent an uncontrolled "driving"
of the total magnetization, we associate to each spin-flip
an opposite flip on a neighboring site. The flips occurr at
each Montecarlo step; since the mean frequency at which the regular
sites are flipped during the simulation is 120 times smaller, 
the resulting effect is to force a zero at the "*" sites.

The precise positions of the sites at which the spin-flips happen
are almost irrelevant; we find in all cases, as appears from the 
following table, a sensible diminution of the correlations between the
spin sums taken along the columns $j_1$ and $j_2$: 

\begin{verbatim}
   Number of "*" sites           p_{12} for L=5
  ===============================================
      3 (Fig. 2.a)                0.19 +/- 0.04
      4 (Fig. 2.b)                0.13 +/- 0.03

\end{verbatim}

It appears therefore that the insertion of variables which 
are driven by an external field damps the correlations in the 
system, and that this mechanism is of a quite general nature, 
although we are not able to give a precise analytical description yet.

\section{Conclusive remarks.}

We have investigated in this paper an unusual interaction mechanism
between gravity and a macroscopic quantum system driven by
external fields. This idea was originally suggested by a possible 
phenomenological application \cite{analysis,pk}, but the
mechanism is interesting also from the purely theoretical point 
of view and deserves further numerical and analytical investigation.

We have shown that under certain conditions the gravitational field 
becomes unstable and may develope singularities, but we have 
not tried to find a physical regularization and to compute the
effect of the regularized singularities yet. Simple physical
analogies show however that they generally reduce the gravitational
long-range correlations. Our next task will be the esplicit estimation 
of the changes in the correlation functions in terms of the squared density
$|\phi_0(x)|^2$ of the Bose condensate and of its squared
gradient $[\partial_\mu \phi_0(x)]^* [\partial^\mu \phi_0(x)]$.


\begin{thebibliography}{99}

\bibitem{veltman}
M.J.G.\ Veltman, in {\it Methods in field theory}, Les Houches
Summer School 1975, ed.s R.\ Balian and J.\ Zinn-Justin,
North-Holland, Amsterdam, 1976.

\bibitem{weinberg}
S.\ Weinberg, Rev.\ Mod.\ Phys.\ {\bf 61} (1989) 1; J.\ Greensite,
Phys.\ Lett.\ {\bf B 291} (1992) 405 and references.

\bibitem{h1}
H.W. Hamber and R.M.\ Williams, Nucl.\ Phys.\ {\bf B 248}
(1984) 392; {\bf B 260} (1985) 747; Phys.\ Lett.\ {\bf
B 157} (1985) 368; Nucl.\ Phys.\ {\bf B 269} (1986) 712.
H.W.\ Hamber, in {\it Les Houches Summer School 1984, Session
XLIII}, North-Holland, Amsterdam, 1986. H.W.\ Hamber,
Phys.\ Rev.\ {\bf D 45} (1992) 507; Nucl.\ Phys.\ {\bf B 400} (1993) 347.

\bibitem{m}
G.\ Modanese, Phys.\ Lett.\ {\bf B 325} (1994) 354; Nucl.\ Phys.\ 
{\bf B 434} (1995) 697; Riv.\ Nuovo Cim.\ {\bf 17}, n.\ 8 (1994).

\bibitem{h2}
H.W.\ Hamber and R.M.\ Williams, Nucl.\ Phys.\ {\bf B 435} (1995) 361.

\bibitem{planck}
L.\ Garay, Int.\ J.\ Math.\ Phys.\ {\bf A 10} (1995) 145.

\bibitem{rovsmo}
C.\ Rovelli, L.\ Smolin, Nucl.\ Phys.\ {\bf B 442} (1995) 593.

\bibitem{graviton}
G.\ Modanese, Phys.\ Lett.\ {\bf B 348} (1995) 51.

\bibitem{tw}
N.C.\ Tsamis and R.P.\ Woodard, Phys.\ Lett.\ {\bf B 301} (1993) 351;
Comm.\ Math.\ Phys.\ {\bf 162} (1994) 217; Ann.\ Phys.\ (N.Y.)
{\bf 238} (1995) 1. 

\bibitem{z}
A.D.\ Dolgov, M.B.\ Einhorn, V.I.\ Zakharov, Phys.\ Rev.\ 
{\bf D 52} (1995) 717.

\bibitem{vdam}
H.\ Van Dam and M.\ Veltman, Nucl.\ Phys.\ {\bf B 22} (1970) 397;
L.H.\ Ford and H.\ Van Dam, Nucl.\ Phys.\ {\bf B 169} (1980) 126.

\bibitem{muz}
I.J.\ Muzinich, S.\ Vokos, Phys.\ Rev.\ {\bf D 52} (1995) 3472.

\bibitem{analysis}
G.\ Modanese, {\it Theoretical analysis of a reported weak
gravitational shielding effect}, report MPI-PhT/95-44, May 1995
(hep-th/9505094); {\it Updating the analysis of Tampere's weak 
gravitational shielding experiment}, report UTF-367/96, Jan 1996
(supr-con/9601001).

\bibitem{pk}
E.\ Podkletnov and R.\ Nieminen, Physica {\bf C 203} (1992) 441;
E.\ Podkletnov and A.D.\ Levit, {\it Gravitational shielding properties
of composite bulk $YBa_2Cu_3O_{7-x}$ superconductor below 70 K under
electro-magnetic field}, Tampere University of Technology report
MSU-chem, January 1995.

\end{thebibliography}
\end{document}